\documentclass[12pt]{iopart}
\usepackage{fleqn,epsfig}
\usepackage{graphicx}
\usepackage{slashed}
\usepackage{bbold}

\def\barpsimux{\overline{\psi}_\mu(x)}

\def\barpsimu{\overline{\psi}_\mu}

\def\onehalf{{1 \over 2}}

\def\gmunu{g_{\mu\nu}}

\def\K {{\cal K}}

\def\ps {\not\!p}
\def\nn{\nonumber }
\def\rf#1{{(\ref{#1})}}
\def\br{\begin{eqnarray}}
\def\er{\end{eqnarray}}

\begin{document}

\title{Higher derivative terms in the $\pi$$\Delta$$N$ interaction: some phenomenological consequences}

\author{\large A. Mariano$^{1,2}$, D. Badagnani$^{1,3}$ and C. Barbero$^{1,2}$}
\address{$^1$Departamento de F\'\i sica,
Facultad de Ciencias Exactas, Universidad Nacional de La Plata,
C.C. 67, 1900 La Plata, Argentina.}
\address{$^2$Instituto de F\'{\i}sica La Plata, CONICET, 1900 La Plata, Argentina.}
\address{$^3$UIDET-CETAD, Facultad de Ingeniería, Universidad Nacional de La Plata}

\ead{daniel@fisica.unlp.edu.ar}

\begin{abstract}

In this paper we implement for the first time the use of a $\pi N \Delta$(1232) vertex interaction
containing both first and second 
order derivative terms, as required by renormalization considerations. 
As as was previously shown both interactions present  formal problems, but can be implemented pertubatively. 
We put an end point to the discussion of which type of interaction 
is the apropriate one: both the usual pi-derivative and the "spin 3/2" gauge invariant that include also a derivative 
in the Delta field should be included in amplitude calculations.
We show that when working within a tree level approach the description of total $\pi N$ scattering in the 
different channels is improved.
\end{abstract}

keywords: DELTA ISOBAR; PION-NUCLEON-ISOBAR INTERACTION; PION-NUCLEON ELASTIC SCATTERING

Pacs: 13.75.-n; 13.75.Gx
\maketitle

\bigskip
\section{Introduction}

For decades, the phenomenology of the $\Delta$(1232 MeV) resonance, was modelled as a Rarita-Schwinger (RS) 
vector spinor field $\Psi_\mu$ with a free Lagrangian

\begin{equation}
{\cal L}_{free}  =  \barpsimux\K(\partial,A)^{\mu\nu}\Psi_{\nu}(x)
\label{eq1}
\end{equation}
where

\begin{eqnarray}
\K(\partial,A)^{\mu\nu} & = & R \left(-\onehalf(1+A)\right)^{\mu\mu'}
                                    \left[ \epsilon_{\mu'\nu'\alpha\beta}\partial^{\alpha}\gamma^{\beta}\gamma_{5} +
                                         im\sigma_{\mu'\nu'}
                                    \right]
                               R \left(-\onehalf(1+A)\right)^{\nu'\nu}
\label{eq2}
\end{eqnarray}
being  $\ensuremath{\sigma_{\mu,\nu}=\frac{i}{2}[\gamma_{\mu},\gamma_{\nu}]=i\gamma_{\mu\nu}},\epsilon_{0123}=1,
\gamma_{5}=i\gamma^{0}\gamma^{1}\gamma^{2}\gamma^{3}$, and
$R_{\mu\nu}(a)=g_{\mu\nu}+a\gamma_{\mu}\gamma_{\nu}$)\footnote{These are Bjorken and Drell conventions.}.
Note that the matrices
$R \left(-\onehalf(1+A)\right)$ appear because by construction the field $\Psi_\mu$ has a spurious spin $1/2$
component  and as a consequence the Lagrangians  ${\cal L}_{free}$  are connected by the contact transformation 
$\Psi^\mu \rightarrow R^{\mu \nu} \Psi_\nu,~~A\rightarrow {A-2a\over 1+4a}$($A\neq -\frac{1}{2}$), 
which change the proportion of the $1/2$ states while leaving the equations of motion invariant.

Interaction
with nucleon $\psi$ and pion $\phi$ fields have been studied using the chiral invariant term which dominates at small energies \cite{Nathetal}:

\begin{equation}
{\cal L}_{I_1} = g_{1} \bar{\psi}\partial_{\mu}\phi^{\dag}\cdot{\bf T} R
                       \left( \onehalf(1+4Z_1)A+Z_1 \right)^{\mu\nu}\Psi_{\nu} + c.c.,
\label{NEK}
\end{equation}
where $I_1$ indicates that we have an interaction with a first derivative, {\it i.e} at order $q_\pi$, and it is also 
invariant (at the level of the equations of motion and the amplitudes) under contact transformations.
It is clear that the A-dependence cancels up in any physical amplitude, whose dependence on $Z_1$ persists.
$Z_1$ is thus a free parameter of the interaction \cite{Marianoetal2012}.

But this interaction has been shown to be problematic from the  formal relativistic field theory point of view
since for certain background pion fields the Fock space becomes non-positive definite, \cite{Hagen}.
However, this term can be used to get
low energy amplitudes since it is the most general  first derivative Lagrangian  respecting 
covariance and chiral symmetry , and admits a nonproblematic perturbative order by order approach for the amplitude
\cite{WeinbergFolk}. Other concern about Delta interaction amplitudes, the existence 
of the so-called ``spin 1/2 background'', has been proved baseless since lowest spin representation contributions are 
present in other cases \cite{Benmerroucheetal}.

In order to try solving these shortcommings of the interaction $I_1$, and based on the supposition that
the key for doing so was the decoupling of the spin 1/2 sector from amplitudes, 
during  the last decade  a new interaction of second derivative order,{\it i.e} $\sim q_\pi q_\Delta$, \cite{Pascalutsa1998}
has been proposed and extensively used. It
has been conjectured to solves both the signature problem and the concern about the spin 1/2 background. It reads

\begin{equation}
{\cal L}_{I2} =
-g_{2}\partial_{\alpha}\bar{{\Psi}}_{\mu}R\left(\frac{1}{2}(1+4Z_2)A+Z_2)\right)_{\,\sigma}^{\mu}\epsilon^{\sigma\nu\alpha\beta}\gamma_{\beta}\gamma_{5}\partial_{\nu}\Phi.
+ c.c.
\label{P}
\end{equation}
It is important to note that this interaction term is the most general second order interaction derivative in the pion
(which is necessary for chiral invariance), provided all free parameters are set such that Lagrange multiplier fields of the
free theory do not aquire dynamics due to the interaction, as explained in the appendix. This criterion, subject to some 
controversy \cite{Benmerroucheetal}, is nonetheless the one used to fix $Z_1=1/2$ for $I_1$ in \cite{Nathetal}, and for $I_2$
it leads to $Z_2=-1/2$, which corresponds to the interaction originally proposed by \cite{Pascalutsa1998}.

Nevertheless we have shown recently that this new interaction presents the same signature problems than the conventional
${\cal L}_{I_1}$
\cite{Badagnanietal2017}, it couples  to a spin 1/2 background in radiative amplitudes and renormalization considerations force
the reintroduction of conventional terms \cite{Badagnanietal2015}.

In addition as expected from general considerations from Effective Field Theory in the resonance region,
(EFT) that consider a contribution of the pion momentum to the power counting of $\delta=(m_N-m_\pi)/\Lambda_{\chi PT}$ or 
$m_\pi/\Lambda_{\chi PT}~\delta^2$ depending of its value, both interactions are of the same order since momentum coming from 
$\partial_\mu \Psi_\nu$ behaves as order $~1$ at threshold \cite{Badagnanietal2015}. 
Furthermore, from the phenomenological point of view, the fit to data from
the new interaction is by no means superior to the conventional \cite{Marianoetal2012} in the resonance region.
All these comments suggest that we should consider $I_1$ and $I_2$ togheter. We consider $I_2$ of higher order in consideration
of the dimmension of the coupling constant and the number of derivatives in the term, in line with \cite{WeinbergNuc}.

The newer interaction $I_2$ could  be easily seen to be simply the next
order in derivatives from the conventional coupling \cite{Badagnanietal2015, Badagnanietal2017}.
The addition of each term $I_k$ implies the incorporation of a new parameter to fit (the corresponding coupling
constant). In this paper we will work with both the $I_1$ and the next derivative order $I_2$,
and will  explore the possible phenomenological consequences of considering its coexistence in the case pion-nucleon scattering.
Then we consider the Lagrangian
\begin{equation}
{\cal L}_{\Delta}  =  {\cal L}_{free} + {\cal L}_{I_1} + {\cal L}_{I_2}.
\label{eq3}
\end{equation}

The structure of the paper is as follows. In Section (2) we build the Feynman amplitude for the elastic\footnote{
In spite we use the terminology ``elastic'' we also will try together the charge exchange $\pi^- p \rightarrow \pi^0 n$
channel with the same approach.} 
$\pi N$ cross section individualizing the cases $g_1=0, g_2=0, g_1,g_2 \neq 0$ in the Lagrangian \rf{eq3}. 
In section (3) we show the numerical results for the $s$ and $u$-channel $\Delta$-exchange contributions to the
total cross section 
and compare the results obtained in both cases for the different $\pi^+ p, \pi^- p, \pi^- p \rightarrow \pi^0 n$ 
channels. In section(4) we summarize our conclusions and in the appendix we show the fixing of the off-shell parameters
$Z_1$ and $Z_2$ and the spin projectors in the RS space.

\section{Elastic pion-nucleon amplitude}

We first calculate the tree level elastic nucleon-pion amplitude from the Lagrangian \rf{eq3}.
Let us first cast the propagator of the RS field in terms of projectors on the different spin sectors. This will allow us
to identify the contributions from each one in the amplitude. First, recall that
$A$ in eq. \rf{eq1} is an arbitrary parameter
while $Z_1$ and $Z_2$ are parameters regulating
the coupling to the off-shell sector $\gamma_\mu \gamma_\nu \Psi^\nu$, as can be seen from the definition of $R(a)$,
and from the fact that on shell we have the constraint $\gamma^\nu \Psi_\nu=0$. 
The fixing for $Z_k$ from theoretical arguments, as we mention above, remains controversial (see
\cite{Benmerroucheetal}) but we stick to the usual criterion used in \cite{Nathetal, Badagnanietal2017} and the appendix
in order to compare our results with the literature.

Using the properties of $R$ matrices in ${\cal L}_{free}$ at \rf{eq3} we can write the general propagator in terms of the propagator for $A=-1$ (which renders
the calculations simpler) as

\begin{eqnarray}
G(p,A)^{\mu\nu} & = & R^{-1}\left(-\onehalf(1+A)\right)_{\alpha}^{\mu}G\left(p,-1\right)^{\alpha\beta}
                       R^{-1}\left(-\onehalf(1+A)\right)_{\beta}^{\nu}
\label{eq4}
\end{eqnarray}
where $G\left(p,-1\right)_{\mu\nu}$ can be put in terms of the well known projectors $P^{3/2}$,
$P_{11,22}^{1/2}$, $P^{1/2}_{21}$ and $P^{1/2}_{12}$ (see the appendix) as

\begin{eqnarray}
G\left(p,-1\right)_{\mu\nu} & =-\left[
                                  \frac{\slashed{p}+m}{p^{2}-m^{2}}P_{\mu\nu}^{3/2}
                                  - \frac{2}{3m^{2}}(\slashed{p}+m)(P_{22}^{1/2})_{\mu\nu}
                                  + \frac{1}{\sqrt{3}m}(P_{12}^{1/2}+P_{21}^{1/2})_{\mu\nu}\right]. &
\label{eq5}
\end{eqnarray}
\begin{figure}
  \begin{center}
  \includegraphics[width=10cm]{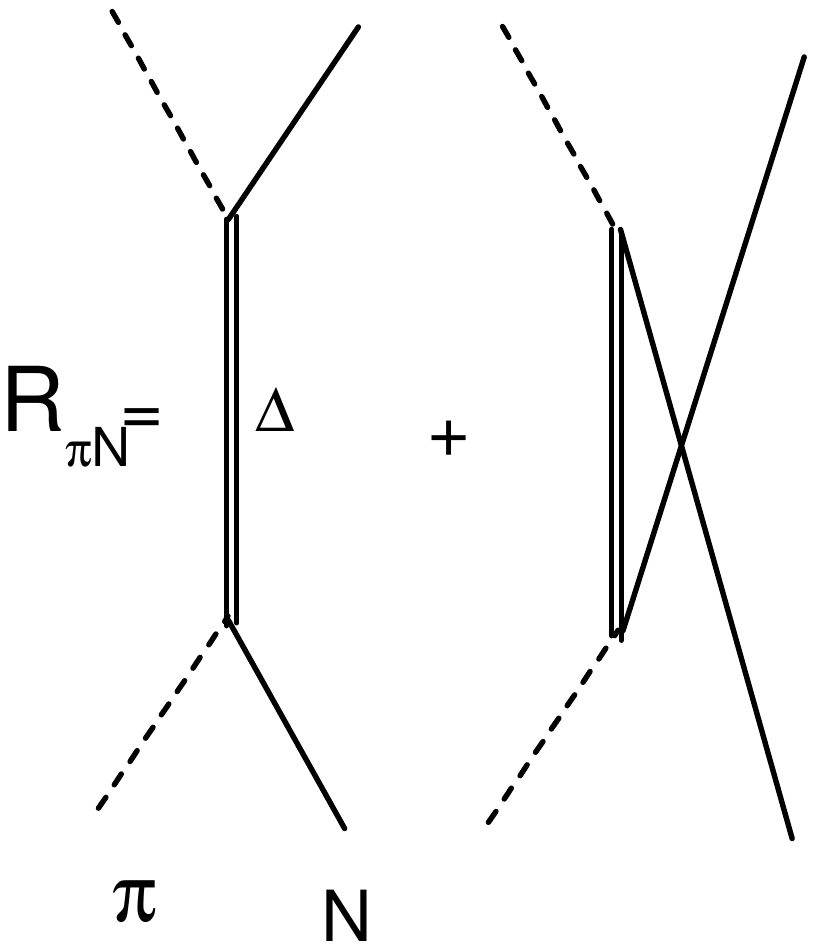}
  \vspace{-8.5cm}
  \caption{Elastic  amplitude of pion ($\pi$) nucleon (N) scattering splited in $s$(left) and $u$ or cross (right) contributions 
  for $\Delta$ resonance.}
  \end{center}
  \label{Fig1}
\end{figure}

Consider the Lagrangian \rf{eq3} with ${\cal L}_{I_1}$ and ${\cal L}_{I_2}$ given by eqs. \rf{NEK} and \rf{P} respectively,
whith $A=-1,Z_1=1/2$, $Z_2=-1/2$; those values for $Z$ were already adopted in refs.\cite{Nathetal} and \cite{Pascalutsa1998}
\begin{eqnarray}
{\cal L}_{\Delta} & = & \barpsimu\left[\epsilon^{\mu\nu\alpha\beta}\partial_{\alpha}\gamma_{\beta}\gamma_{5}
                         +(im\sigma^{\mu\nu}\equiv mR(-1)^{\mu\nu})\right]\Psi_{\nu}  \nonumber  \\
                  &   & + g_{1}\bar{\psi}\partial_{\mu}\phi^{\dag}\cdot{\bf T}R(-1)^{\mu\nu}
                         \Psi_{\nu}+g_{1}\bar{\Psi}_{\mu}R\left(-1\right)^{\mu\nu}\partial_{\nu}
                         \phi\cdot{\bf T^{\dag}}\psi   \nonumber  \\
                  &   & - g_{2}\,\bar{\psi}\partial_{\mu}\phi^{\dag}\cdot{\bf T}
                          \epsilon^{\mu\nu\alpha\beta}\partial_{\alpha}\gamma_{\beta}\gamma_{5}\Psi_{\nu}
                          +g_{2}\partial_{\alpha}\barpsimu\epsilon^{\mu\nu\alpha\beta}\gamma_{\beta}\gamma_{5}
                          \partial_{\nu}\phi\cdot{\bf T^{\dag}}\psi,  \label{eq19}
\end{eqnarray}
where ${\bf T}$ are the $N\rightarrow \Delta$ isospin excitation operators. Now from it and the  propagator \rf{eq5} we can 
calculate the resonance elastic amplitude contribution($R_{\pi N}$) shown in the fig.\rf{Fig1}.
Omitting the nucleon spinors and isospin factors,
letting the incoming and outgoing pions momentum be noted $k$ and $k'$ respectively  and the $\Delta$ momentum 
by $p$ the {\it s}-channel amplitude reads

\begin{eqnarray}
R_{\pi N} & = &   g_{1}^{2}k'_{\mu}R^{\mu\alpha}G_{\alpha\beta}(p)R^{\beta\nu}k_{\nu}
                 +g_{2}^{2}k'_{\mu}(-i)\Gamma^{\mu\alpha}(p)G_{\alpha\beta}(p)(-i)\Gamma^{\beta\nu}(p)k_{\nu} \nonumber  \\
          &   &  -g_{1}g_{2}  k'_{\mu}R^{\mu\alpha}G_{\alpha\beta}(p)(-i)\Gamma^{\beta\nu}(p)k_{\nu}
                          -g_{1}g_{2}k'_{\mu}(-i)\Gamma(p)^{\mu\alpha}G_{\alpha\beta}(p)R^{\beta\nu}k_{\nu} \label{eq20}
\end{eqnarray}
where $\Gamma^{\mu\nu} (p) = \epsilon^{\mu\nu\alpha\beta}\gamma_{\beta}\gamma_{5}p_{\alpha}$, $R\equiv R(-1)$
and where we have used
$\partial_{\mu}\psi,\phi,\Psi_{\nu}\sim-iq_{\mu}\psi,\phi,\Psi_{\nu}$. The {\it u}(cross) channel contribution  is 
obtained by simply replacing
$p$ by $p-k-k'$ and $k$ by $k'$ in the former expression. Observe that the first two terms correspond to 
the first and second order (in derivatives) contributions to the Lagrangian
respectively while the last two can be construed as interference terms between them. Let us analyze first
the amplitude for the first order  derivative interactions. If $g_2=0$ we get the {\it s}-chanel amplitude

\bigskip

\begin{eqnarray}
R^{g_2=0}_{\pi N}
       &=& -(g_1)^2 \frac{\slashed{p}+m}{p^2-m^2} P^{3/2}_{\mu \nu} k_\mu k'_\nu \nonumber \\
       & & - \frac{(g_1)^2}{m^2}
          \left[
             2(\slashed{p}+m) P^{1/2}_{11 \; \mu \nu} + m\sqrt{3} (P^{1/2}_{12}+P^{1/2}_{21})_{\mu \nu}
          \right] k_\mu k'_\nu,
\label{ampleading}
\end{eqnarray}
where the relations
\begin{eqnarray}
R^{\mu\nu} & = &P_{3/2}^{\mu\nu}-\sqrt{3}(P_{12}^{1/2}+P_{21}^{1/2})^{\mu\nu} -2(P_{11}^{1/2})^{\mu\nu},\nn\\
(p^{\mu}\gamma^{\nu}-\gamma^{\mu}p^{\nu}) & = & \sqrt{3}\ps(P_{12}^{1/2}+P_{21}^{1/2})^{\mu\nu}=-\sqrt{3}(P_{12}^{1/2}+P_{21}^{1/2})^{\mu\nu}\ps,\nn\\
\left[P^{3/2}_{\mu \nu},\ps\right]&=&\gamma^\mu P^{3/2}_{\mu\nu}=P^{3/2}_{\mu\nu}\gamma^{\nu}=0,\label{gamma0}
\label{gamma0}
\end{eqnarray}
were used.
The second term in \rf{ampleading} is the so-called ``spin 1/2 background'', but what is relevant for the asymptotic behavior
of the amplitude is that it represents a non-pole (without a pole)  contribution which grows with $p$, since the projectors 
go as $p^0$ (see Appendix). On the other hand, if
$g_1=0$ the second order derivative amplitude can be expressed also as a pole and non-pole  term:
\begin{eqnarray}
R^{g_1=0}_{\pi N}
       &=& -(g_2)^2 m^2 \frac{\slashed{p}+m}{p^2-m^2} P^{3/2}_{\mu \nu} k_\mu k'_\nu \nonumber \\
       & & - (g_2)^2  (\slashed{p}+m) P^{3/2}_{\mu \nu} k_\mu k'_\nu
\label{ampderivative}
\end{eqnarray}
where we have used eqs.\rf{gamma0} and that
\br
-i\Gamma(p)^{\mu\nu} & = & -\ps R^{\mu\nu}-(p^{\mu}\gamma^{\nu}-\gamma^{\mu}p^{\nu})=-R^{\mu\nu}\ps+(p^{\mu}\gamma^{\nu}-\gamma^{\mu}p^{\nu}).\label{gamma1}
\er
Observe that, except for the dimensions of the coupling constants, the pole terms for both amplitudes are identical in form.
Additionaly there is also a non-pole background term as badly behaved asymptotically as the non-pole  term in
\rf{ampleading}. We well might call it a ``spin 3/2 background''.
Putting now both interactions together the amplitude reads

\begin{eqnarray}
R_{\pi N} & = & \left({g_{1}+mg_{2}\over m}\right)^{2}k'_{\mu}\left\{ -p^{2}\frac{\ps+m}{p^{2}-m^{2}}P^{\mu\nu}_{3/2}\right\} k_{\nu}\nn\\
&+&\left[\left({g_{1}+mg_{2}\over m}\right)^{2}-g_{2}^{2}\right]k'_{\mu}\left\{ (\ps+m)R^{\mu\nu}+(p^{\mu}\gamma^{\nu}-\gamma^{\mu}p^{\nu})\right\} k_{\nu}\nn\\
 & - & 2g_{1}g_{2}k'_{\mu}R^{\mu\nu}k_\nu,\label{bothint1}\\
 & = & -\left(g_{1}+mg_{2}\right)^{2}k'_{\mu}\left\{ \frac{\ps+m}{p^{2}-m^{2}}P^{\mu\nu}_{3/2}\right\} k_{\nu}\nn\\&-&\left({g_{1}+mg_{2}\over m}\right)^{2}k'_{\mu}\left\{2(\ps+m)(P_{11}^{1/2})^{\mu\nu})+m\sqrt{3}(P_{12}^{1/2}+P_{21}^{1/2})^{\mu\nu} \right\} k_{\nu}\nn\\
&+&g_{2}^{2}k'_{\mu}\left\{2(\ps+m)(P_{11}^{1/2})^{\mu\nu})+m\sqrt{3}(P_{12}^{1/2}+P_{21}^{1/2})^{\mu\nu} \right\} k_{\nu}\nn\\
&-& g_2^2 k'_\mu (\ps+m)P_{3/2}^{\mu\nu}k_\nu \nn\\
 &-& 2g_1g_2k'_\mu\left(P_{3/2}^{\mu\nu}-2(P_{11}^{1/2})^{\mu\nu}-\sqrt{3}(P_{12}^{1/2}+P_{21}^{1/2})^{\mu\nu}  \right)k_\nu.\nn\\
&=&  -\left({g_{1}+mg_{2}\over m}\right)^{2}k'_{\mu}\left\{ \frac{\ps+m}{p^{2}-m^{2}}P^{\mu\nu}_{3/2}\right\} k_{\nu}\nn\\
&-&{g_{1}^{2}\over m^2}k'_{\mu}\left\{2(\ps+m)(P_{11}^{1/2})^{\mu\nu})+m\sqrt{3}(P_{12}^{1/2}+P_{21}^{1/2})^{\mu\nu} \right\} k_{\nu}\nn\\
&-& g_2^2 k'_\mu (\ps+m)P_{3/2}^{\mu\nu}k_\nu\nn\\
&-& {2g_1g_2\over m} k'_\mu \left(2\ps (P^{1/2}_{11})^{\mu\nu}+mP_{3/2}^{\mu\nu}\right) k_\nu,
\label{bothint2}
\end{eqnarray}
where we have assumed that the projectors are defined for each $p^2$, that is the case for the $s$-channel contribution 
for which $p^2>0$ always, while for the $u$-channel one, since $p^2$ could be arbitrary small, 
it is preferable to express the amplitude without separating pole from non-pole terms, as in eq.\rf{bothint1}.
As can be seen in eq. \rf{bothint2}, the first term corresponds to the pole contribution both in \rf{ampleading} 
with $g_2=0$  and  \rf{ampderivative} with $g_1=0$, but now with a coupling constant $g=(g_1+m g_2)$. 
The third and fourth terms are the corresponding backgrounds from \rf{ampleading} and \rf{ampderivative},  
the last term being a background contribution coming from the interference of both vertices.
If we use these interactions in the region of the resonance, of
course, it is unimportant the individual values of $g_1$ and $g_2$, only the combination $g=(g_1+m g_2)$ would be observable, 
but for higher values the backgrounds become relevant, and we might ask if by a judiciously adjustment of both coupling 
constants the bad high
energy behavior (generated by the non-pole background) of the amplitude can be moderated. 
Let us write the complete amplitude for the $s$-chanel in terms of $g$ and a parameter
$\kappa$ assuming that $g_1 + m g_2 = g $, in order to keep the old peak adjustment, and $g_2 = \kappa g/m$. In this way

\begin{eqnarray}
g_1     & = & (1-\kappa) g \\
g_2     & = & \kappa{g\over m} 
\end{eqnarray}
Observe that for $\kappa=0$ we obtain the amplitude for pure leading first derivative interaction, while in the limit $\kappa =1$
we get pure second derivative interaction. The amplitude reads

\begin{eqnarray}
R_{\pi N}
&=&  - g^2 k'_{\mu}\left\{ \frac{\ps+m}{p^{2}-m^{2}}P^{\mu\nu}_{3/2}\right\} k_{\nu}\nn\\
&-&{(1-\kappa)^2\over m^2}g^2 k'_{\mu}\left\{2(\ps+m)(P_{11}^{1/2})^{\mu\nu})+m\sqrt{3}(P_{12}^{1/2}+P_{21}^{1/2})^{\mu\nu} \right\} k_{\nu}\nn\\
&-& {\kappa^2\over m^2} g^2 k'_\mu (\ps+m)P_{3/2}^{\mu\nu}k_\nu \nn\\
&-& {2(1-\kappa)\kappa\over m^2}g^2 \left(2\ps P^{1/2}_{11}+mP_{3/2}\right)^{\mu\nu}.
\label{bothint2}
\end{eqnarray}
Observe that we get the same peak contribution (if we assume the same value of $g\equiv {f_{\pi N\Delta}\over m_\pi}$ 
used when $g_2=0, g_1=g$ but as $\kappa,(1-\kappa) < 1$ the backgrounds are reduced by a smaller factor $\kappa^2,(1-\kappa)^2$, 
and with an interference background (last term) coming from the last term in eq.\rf{bothint2} of the same order 
but enlarged by a factor 2. Nevertheless, as we will see, the $P^{1/2}_{11},P_{3/2}$ contributions present in this 
term are small and also the $1/2$ backgrounds do not interfere with the peak $3/2$ amplitude while the $3/2$ one does.

\begin{figure}
\begin{center}
  \includegraphics[width=10cm]{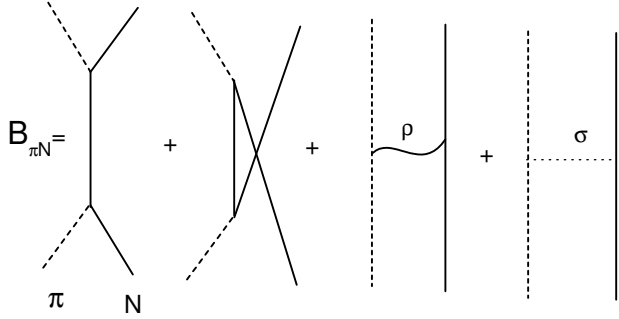}
  \vspace{-7cm}
  \caption{Background non-resonant contributions to the $\pi N$ amplitude.}
  \label{Fig2}
  \end{center}
\end{figure}

\section{Numeric calculations}

Now we are going to calculate the resonance $s$ and $u$-channel contributions shown in fig. \rf{Fig1} to the  $\pi N$ 
total elastic cross section. In addition, to reproduce experimental data we must include other backgrounds 
$(B_{\pi N})$ to the amplitude as shown in fig. \rf{Fig2} (see \cite{Lopez02} for details).
As the denominator of the pole term in eq.\rf{bothint2} is not defined at $p^2=m^2$ for the s amplitude we 
should include the unstable character of the $\Delta$. The simplest way to do it is the complex mass scheme (CMS) 
where we make the replacement $m\rightarrow m+i\Gamma$ in the {\it full} propagator, being $\Gamma$ the $\Delta$ 
width \cite{Lopez02}. A more accurate procedure would be the use of the energy dependent $\Delta$ self-energy 
taking into account the $\Delta $ mixing with $\pi N$ states at one or higher loop bubbles 
to all order\cite{Barbero12}. Nevertheless, since we are interested in the qualitative behavior of the different 
components of the $\pi N\Delta$ interaction, we use the simpler CMS scheme. 
Within this approach and with the $I_1$ interaction, $m$, $\Gamma$ and $g$  were fitted to describe the $\pi^+p$ 
total elastic cross section satisfactorily \cite{Lopez02} in the resonance region, giving 
$m=1211.2 MeV, \Gamma=88.16 MeV, g=0.316$.
As was shown previously \cite{Marianoetal2012}, results with both interactions separately do not change appreciably 
the prediction of the total elastic cross section
in the resonance region, the $I_1$ interaction fitting the experimental data slightly better. For this reason we will fix 
$g_1 + m g_2 = g =0.316$ with the $m$ and $\Gamma$ values above reported. 
Nevertheless we want to analyze the cross section for energies above the resonance since in experiments the strong $\pi N$ 
interaction is tipically probed for $\pi N$ invariant masses up to $2 \; GeV$ in pion-photoproduction and weak-production, 
using models generalizing the CMS to energy-dependent width but the same $B_{\pi N}$ background \cite{Hernandezetal2007}. 
In addition, there are channels for which the $\Delta$ $s$-channel contribution is suppressed by the isospin 
factors with respect to the $\Delta$ $u$-channel one, as in $\pi^-p$ scattering and then we want to analyze also 
the effect of using an interaction of the form $I_1 + I_2$ for this case.

\begin{figure}\begin{center}
  \includegraphics[width=12cm]{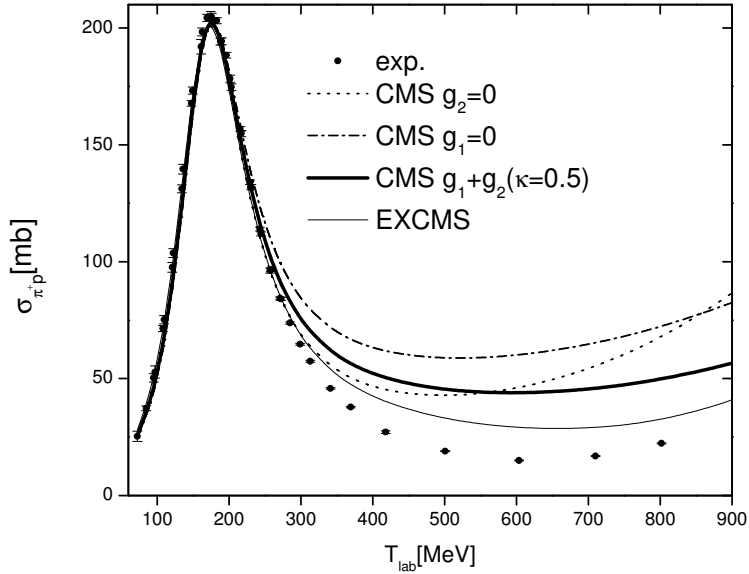}
  \vspace{-6cm}
  \caption{Elastic $\pi^+N$ cross section within the different approaches described in the text. Data from \cite{GWDC}.}
  \label{Fig3}
  \end{center}
\end{figure}

Firstly we fix $\kappa$ in eq. \rf{bothint2} in order to get the closer approach to the $\pi^+ p$ elastic cross 
section experimental data, which were used previously for fixing the $\Delta(1232)$  parameters in the resonance  region. 
The amplitude  is built with the diagrams in figs. \rf{Fig1} and \rf{Fig2}, {\it i.e} $M_{\pi N} = R_{\pi N} + B_{\pi N}$, 
and reach pion kinetic energies ($T_{lab}$) of the order of 900 MeV ($W_{\pi N}\sim 2$ GeV). 
We find that for $\kappa=1/2$ the model is closest to the data. 
In fig. \rf{Fig3}, we show the results for the total elastic 
cross section for this value for $\kappa$ together with the results for $\kappa =0$ (corresponding to $g_2=0$) 
and $\kappa=1$ ($g_1=0$)
to appreciate the improvement achieved. Also we show the results within the so called EXCMS approximation 
with $g_2=0$, obtained in a previous work by solving the Bethe-Salpeter equation for the $\Delta$ propagator 
in presence of a one-loop self energy and replacing at the end the renormalized mass as in the CMS approach 
\cite{Barbero15}, but still keeping the tree label approximation for the background 
$B_{\pi N}$\footnote{A better description would be got by introducing the rescattering of the background, but this at 
the price of introducing form factors to regularize the integrals involved for intermediate pion-nucleon states.}. 
As can be observed, by no means within the CMS scheme the $I_2$ interaction gets a better description than the $I_1$ one, and 
the use of $I_1+I_2$  improves the individual descriptions maintaining the right behavior in the pole peak.

In order to understand how both interactions cooperate to improve the high energy behavior of the amplitude
we are going to separate the resonance peak and background amplitude contributions, which only makes sense for the 
$s$-channel $\Delta$ interchange, since for the $u$-channel the pole $3/2$ term in eq. \rf{bothint2} alone is ill defined
due to the possible vanishing of $p^2$. We begin with the $s$-channel: in fig. \rf{Fig4} we show the pole contribution 
from the resonant amplitude $R_{\pi N}(pole)$ (first term in \rf{bothint2}) together with the background 
$R_{\pi N}(1/2)$ (second term) contribution when $\kappa=0$ ($g_2=0$) which, as can be seen, is the largest term
at high energies. We show also the $R_{\pi N}(3/2)$ (third term) background contribution when $\kappa=1$ ($g_1=0$),
and finally we separate $2\ps P_{11}$ and $m P_{3/2}$ contributions in order to analyze the size of the last term in 
\rf{bothint2}. Finally the full background for the $I_1+I_2$ case $R_{\pi N}(1/2+3/2)$ when $\kappa=0.5$ 
(second + third + fourth terms) is shown.
\begin{figure}\begin{center}
  \includegraphics[width=14cm]{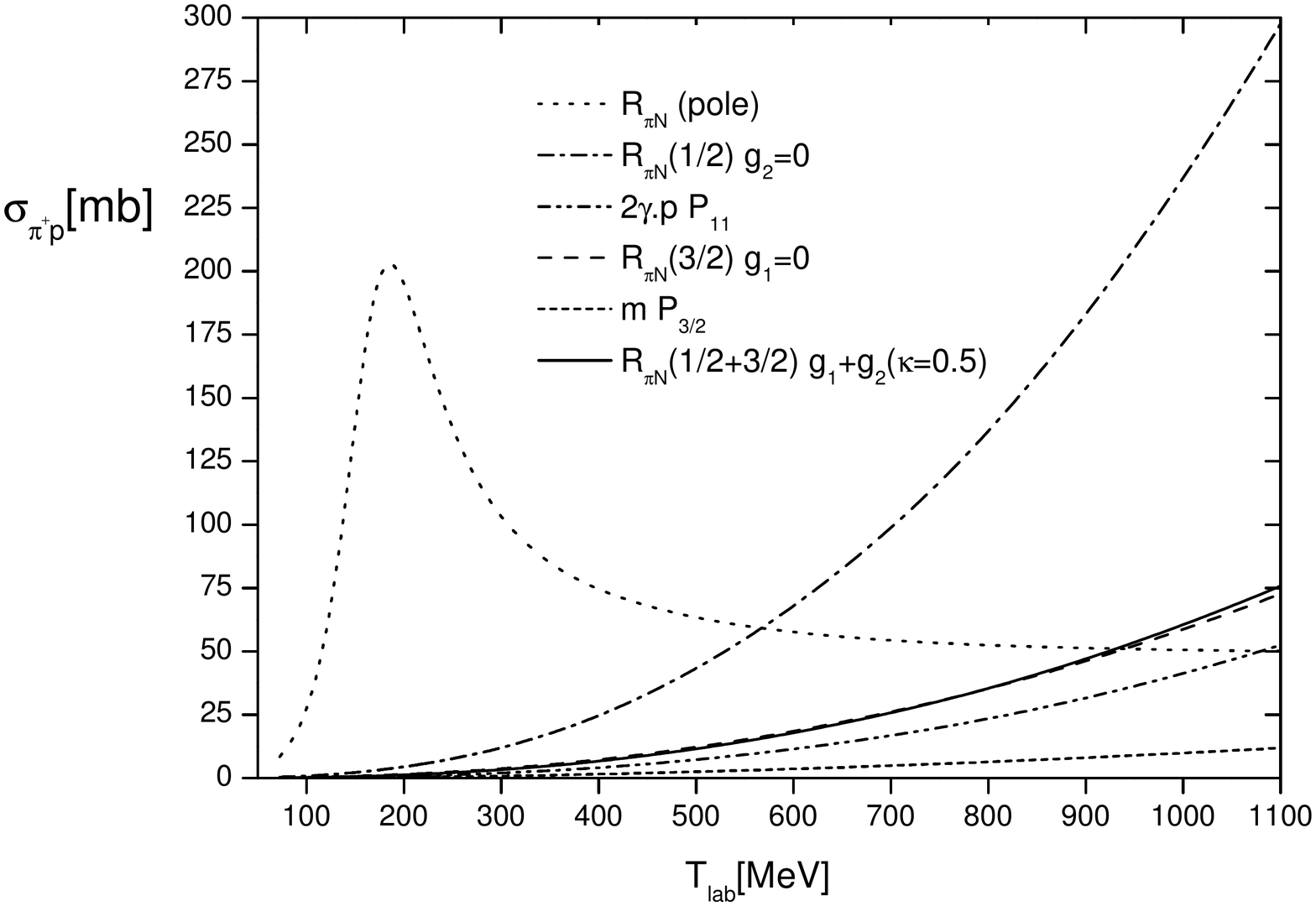}
  \vspace{-1cm}
  \caption{Contributions to the s-channel from the $R_{\pi N}(pole)$ together with different contributions from background to $R_{\pi N}$ as indicated. We slightly extend the energy range to individualize better each contribution }
  \label{Fig4}
  \end{center}
\end{figure}

As can be seen, the second  term in eq. \rf{bothint2} is strongly supressed since its contribution is now
weighted by a factor $1/16$ in the 
cross section for $\kappa=1/2$. Nevertheless, since it is larger than other contributions, it still contributes
roughly a  $25\%$ of the $\kappa=1$ $R_{\pi N}(3/2)$ background at $T_{lab}=1100 /; MeV$, which we will use as a reference.
Also the third contribution $R_{\pi N}(3/2)$ is reduced to $6.25\%$ of our reference, making an even 
smaller contribution to the 
background.
It might be surprising the simmilarity between the backgrounds $R_{\pi N}$ for $g_1=0$ and $\kappa=0.5$, since they
have clearly different origins in spin content. This simmilarity can be understood if we observe that we are
plotting total (i.e. integrated) cross sections, since it can be shown that 
$\sum_{spins}\int d\theta |\bar{u}~2 \slashed{p} P_{11}^{\mu\nu}q'_\mu q_\nu u |^2 =
2\sum_{spins}\int d\theta |\bar{u} \slashed{p} P_{3/2}^{\mu\nu}u q'_\mu q_\nu u |^2$.
The effect of the interference between the $R_{\pi N}(pole)$ and $R_{\pi N}(1/2+3/2)$ s-amplitudes to 
the total cross section is shown in fig.\rf{Fig5}.
\begin{figure}\begin{center}
  \includegraphics[width=12cm]{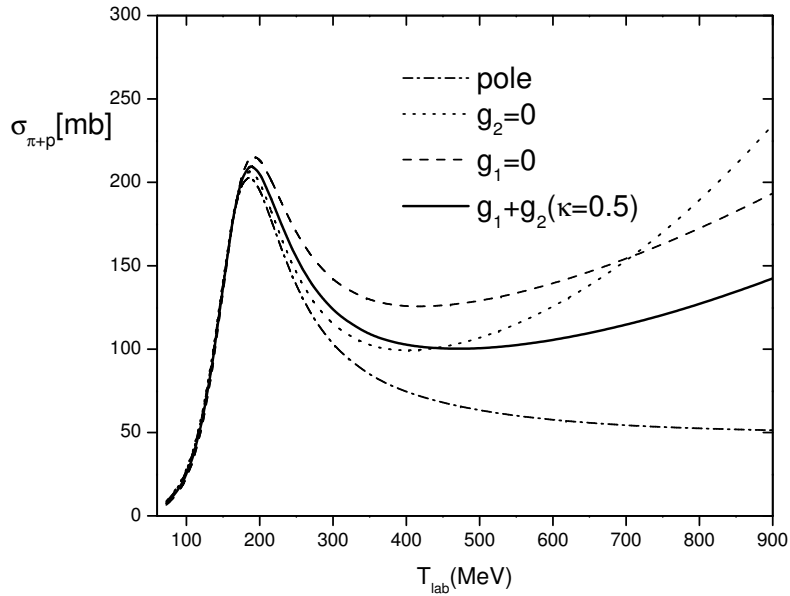}
\vspace{-0.5cm}
  \caption{$R_{\pi N}=R_{\pi N}(pole)+R_{\pi N}(1/2+3/2)$ contribution to the cross section.}
  \label{Fig5}
  \end{center}
\end{figure}

The improvement due to the weighting factors in eq. \rf{bothint2}
is clear, but it gets still better when considering the interference between $R_{\pi N}$ and $B_{\pi N}$
as shown in fig. \rf{Fig3}.

It is also important to analyze all the mentioned effects on the $u$-channel (second term in fig.\rf{Fig1}), 
which amplitude we call $R_{\pi N}(cross)$ in fig. \rf{Fig6} and where now the different contributions are not separated 
in spin content since, as mentioned earlier, that separation leads to ill defined terms.
In fig.\rf{Fig6} we show the $u$-channel contribution to the total cross section in the $\pi^+ p$ scattering together with the
corresponding contributions if $g_2=0$ and if $g_1=0$. We see that for $I_1+I_2$ the contribution lies between them.
Observe that the cross section in the case $g_1=0$ is quite suppressed due to the fact that $p^2<0$ in the resonance region.
This result represents an improvement, 
as can be seen when we calculate the elastic $\pi^-p$ cross section where 
the $u$-channel contribution is the most important while the $s$-channel one is suppressed by the isospin factors. 
In the fig.\rf{Fig7} we compare our calculation with the different interactions to the elastic $\pi^-p$ cross section. 
Here can also be seen the improvement when using both interactions together, in spite that in this case for higher energies we 
have more excited resonances than the $\Delta$ that is the only included in or model. 
This is so because the $\pi^-p$ state has an stronger isospin $1/2$ component, with many $1/2$ isospin resonances 
in the region.

\begin{figure}\begin{center}
  \includegraphics[width=12cm]{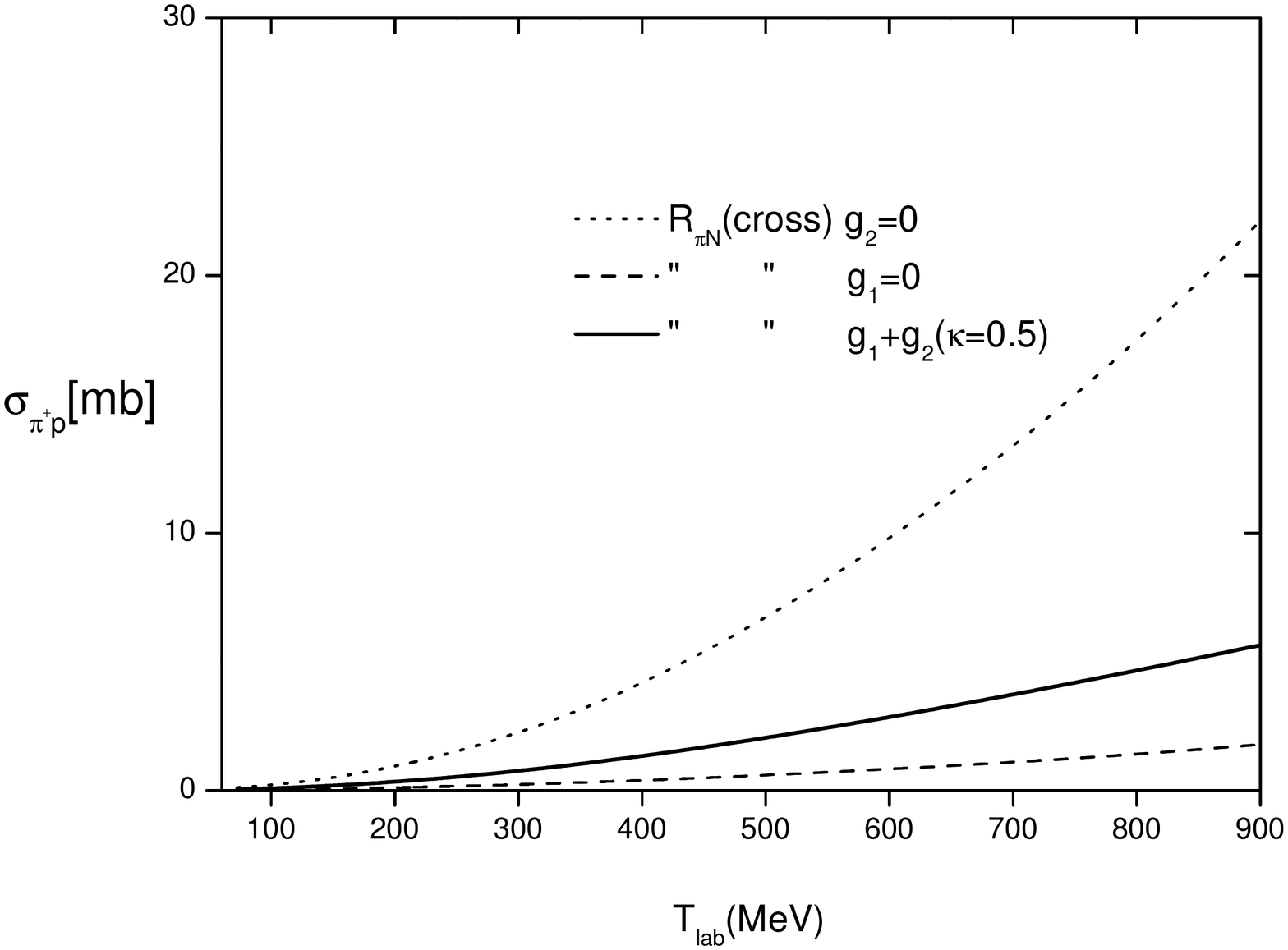}
\vspace{-0.5cm}
  \caption{$R_{\pi N}(cross)$ term  contribution to the total elastic cross section.}
  \label{Fig6}
  \end{center}
\end{figure}
\begin{figure}\begin{center}
  \includegraphics[width=12cm]{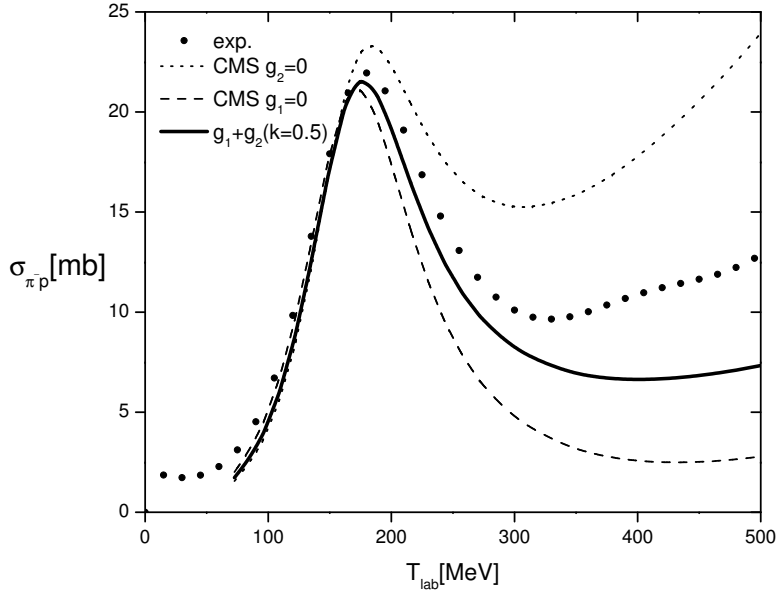}
\vspace{-1cm}
  \caption{Elastic $\pi^- p$ cross section.}
  \label{Fig7}
  \end{center}
\end{figure}
\begin{figure}\begin{center}
  \includegraphics[width=12cm]{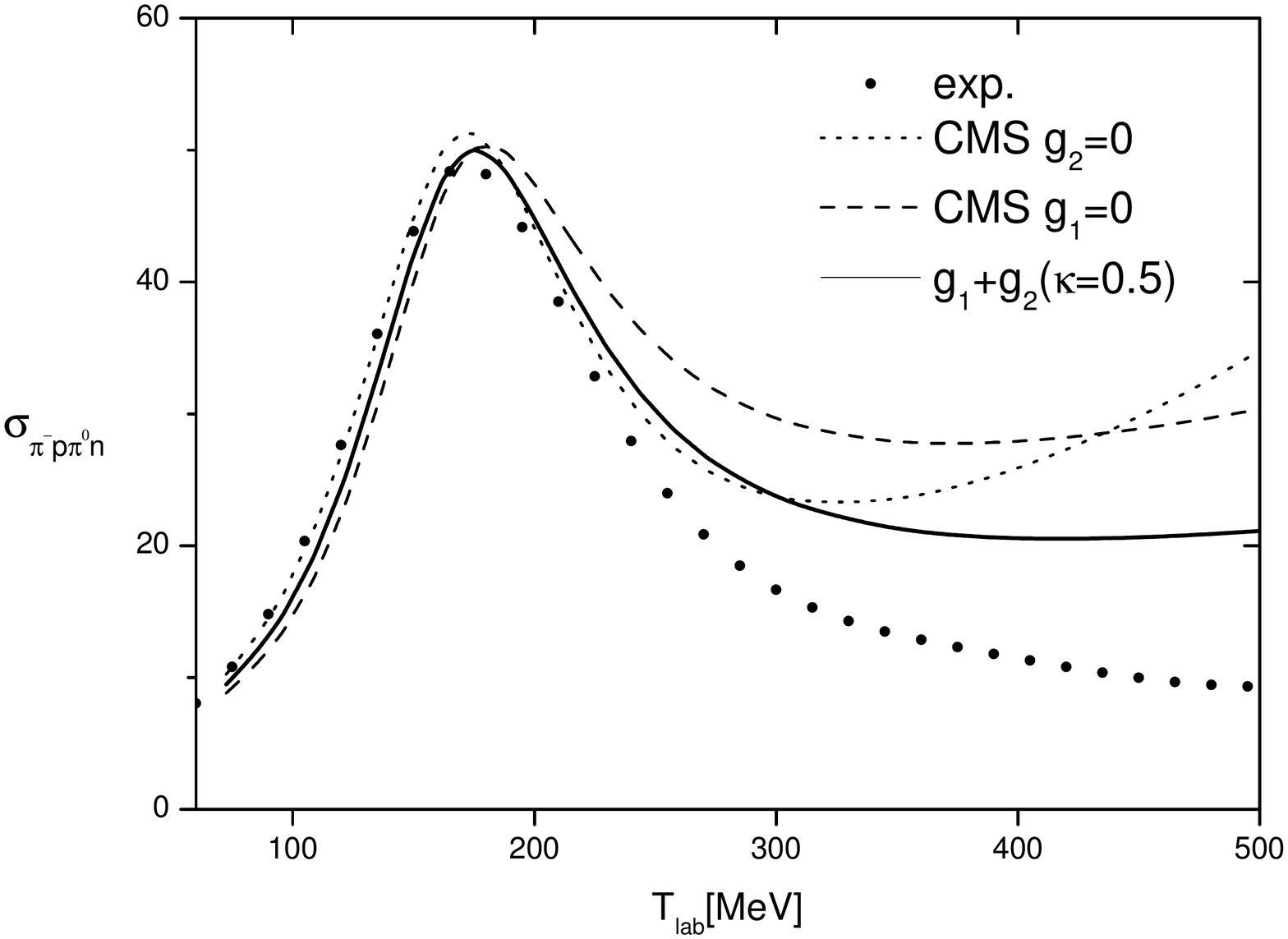}
\vspace{-1cm}
  \caption{$\pi^- p \rightarrow \pi^0 n$ cross section.}
  \label{Fig8}
  \end{center}
\end{figure}

Finally, and for the sake of completeness, we show in fig.\rf{Fig8} the same results for the charge 
exchange $\pi^- p \rightarrow \pi^0 n$ channel. Here, the improvement using $I_1+I_2$ is moderate since in this case 
the amplitude is built as $\sqrt{2}/3 (s-u)$ contributions, thus the overvaluation at high energies in the $\pi^+p$ channel 
(mainly $s$-like) and the undervaluation in the $\pi^-p$ one (mainly $u$-like), leads to a worse behavior for this channel. 
In spite of this, using $I_1+I_2$ is represents an improvement.

\section{Conclusions}

To date, publications devoted to the description of pion-production reactions describe the $\pi N\Delta$ vertex by either 
using the conventional first derivative pion field Lagrangian eq.\rf{NEK}  or the more recent second order derivative 
Lagrangiam in eq.\rf{P}, as if they were mutually excluding possibilities.
As mentioned earlier, both interaction vertexes have problems from the formal point of view, but can be used in a 
perturbative approach. We have shown that, within the spirit of Effective Field Theories, and especially what
Weinberg called ``folk theorem'', in order to reproduce low energy amplitudes these interaction terms are the first two
of an infinite series respecting the point invariance $\Psi_\mu \rightarrow \Psi_\mu + \gamma_\mu \gamma^\nu \Psi_\nu$,
which should provide increasing levels of accuracy.
We show here that using both interaction terms together the fit to experimental data above the resonance region
improves in all channels. 
Results for the total elastic cross section in the $\pi^+p$ and $\pi^- p$ channels and also for the 
$\pi^- p\rightarrow \pi^0n$ inelastic one, are improved when we work in a tree level approach within a complex mass scheme
for the $\Delta$ resonance. This is physically sound, since we expect that, in order to reach higher energies, higher
orders of a $p^2$ expansion of the interaction should be needed.
Of course the description would be additionally improved by solving the Bethe-Salpeter equation for the $\Delta$ propagator 
using both interaction vertexes to account for the mixing of the $\Delta$ with $\pi$ and $N$ states, or by including 
rescattering of the non-resonance background contributions, but we will analyze this in a future publication.
Without this it make no sense qualify our results trough a $\chi^2$ value.

\section{Appendix}

\subsection{Free parameters in the $I_2$ interaction}

To show the generality of the form of ${\cal L}_{I_2}$ observe that the general covariant form of the coefficients to
contract with pion and a second derivative is,
taking for simplicity $A=-1$ (see section 2), omitting isospin factors and  integrating by parts
\br
{\cal L}_{I2}(A=-1) = -g_2\bar{{\Psi}}_{\mu}R(-{1 \over 2}-Z_2)^\mu_\sigma{\cal M}^{\sigma\alpha\nu}\partial_\alpha(\partial_{\nu}\Phi\psi)+c.c. ,\label{5b}
\er
where the most general tensor structure  for ${\cal M}$ is 
${\cal M}^{\mu\alpha\nu}= Z_1 \gamma^\mu \gamma^\alpha \gamma^\nu +
Z_2 g^{\mu\alpha}\gamma^\nu+z_2 g^{\mu\nu}\gamma^\alpha+z_3 g^{\alpha\nu}\gamma^\mu$.
Observe that in the free RS lagrangian in \rf{eq1} and \rf{eq2}, there is no term
containing $\dot{\Psi}^0$ for $A=-1$. So, the equation of motion for it is a true constraint, and $\Psi^0$ has no dynamics. 
It is necessary then that interactions
do not change that (see ref.\cite{Badagnanietal2017} appendix A). But since ${\cal L}_{I_2}$ contributes a $\dot{\Psi}^0$
in the equation of motion for $\psi$ via $R(-{1 \over 2}-Z_2)^0_\sigma{\cal M}^{\sigma\alpha\nu}$, 
the condition for $\dot{\Psi}^0$ not appearing in the equations of motion is that this contribution contains no time 
derivative of any of the other fields of the theory. 
This can be realized if $R$ is diagonal achieved with  $Z_2=-1/2$, and if ${\cal M}^{0\alpha0}={\cal M}^{00\nu}=0$. 
This leads to
\br
{\cal M}^{\mu\alpha\nu}=Z_1[\gamma^{\mu}\gamma^{\alpha}\gamma^{\nu}+g^{\mu\nu}\gamma^\alpha-g^{\mu\alpha}\gamma^\nu-g^{\alpha\nu}\gamma^\mu]
=i\epsilon^{\mu\nu\alpha\rho}\gamma_\rho\gamma_5,\label{5c}
\er
where we have used a property of gamma matrices. Finally if we replace eq.\rf{5c} in \rf{5b} we get eq.\rf{P} for $A=-1$.

\subsection{Spin projectors}

We have introduced $P^k_{ij}$  which projects on the $k=3/2$, $1/2$ sector of the representation space,
with $i,j=1,2$ indicating the subsectors of the $1/2$ subspace, and are defined as

\br (P^{3/2})_{\mu \nu}&=&g_{\mu \nu}-{ 1 \over 3}\gamma_\mu\gamma_\nu -{1\over 3p^2}\left[\ps \gamma_\mu p_\nu + p_\mu \gamma_\nu\ps\right],\nonumber\\
(P^{1/2}_{22})_{\mu \nu}&=&{p_\mu p_\nu \over p^2},\nonumber\\
(P^{1/2}_{11})_{\mu \nu}&=& \gmunu - P^{3/2}_{\mu \nu}
-(P^{1/2}_{22})_{\mu \nu},\nn\\
&=&
(g_{\mu \alpha}-{p_\mu  p_\alpha\over p^2})(1/3\gamma^\alpha \gamma^\beta)(g_{\beta \nu}-{p_\beta p_\nu \over p^2}),\nonumber\\
(P^{1/2}_{12})_{\mu \nu}&=&{1\over \sqrt{3}p^2}(p_\mu p_\nu -\ps \gamma_\mu p_\nu),\nonumber\\
(P^{1/2}_{21})_{\mu \nu}&=&{1\over \sqrt{3}p^2}(-p_\mu
p_\nu+\ps p_\mu \gamma_\nu ).\label{spinprojectors}
\er

\
\section{Acknowledgments}
A. Mariano and C. Barbero fellows to CONICET and UNLP. D. Badagnani fellow to UNLP.

\end{document}